\documentclass[
 reprint,
superscriptaddress,
showpacs,preprintnumbers,
 amsmath,amssymb,
 aps,
 prl,
 lengthcheck,%
]{revtex4-1}

\usepackage{graphicx}% Include figure files
\usepackage{dcolumn}% Align table columns on decimal point
\usepackage{bm}% bold math
\usepackage{hyperref}% add hypertext capabilities
\usepackage{color}

\begin{document}

%\preprint{APS/123-QED}

\title{Unusual Shubnikov-de Haas oscillations in BiTeCl}

\author{C. Martin}
\affiliation{Ramapo College of New Jersey, Mahwah, NJ, 07430, USA}
\author{A. V. Suslov}
\affiliation{National High Magnetic Field Laboratory, Tallahassee, FL, 32310, USA}
\author{S. Buvaev}
\affiliation{Department of Physics, University of Florida, Gainesville, Florida 32611, USA}
\author{A.F. Hebard}
\affiliation{Department of Physics, University of Florida, Gainesville, Florida 32611, USA}
\author{P. Bugnon}
\affiliation{Institute of Physics of Complex Matter, Ecole Polytechnique Federal de Lausanne, CH-1015 Lausanne, Switzerland}
\author{H. Berger}
\affiliation{Institute of Physics of Complex Matter, Ecole Polytechnique Federal de Lausanne, CH-1015 Lausanne, Switzerland}
\author{A. Magrez}
\affiliation{Institute of Physics of Complex Matter, Ecole Polytechnique Federal de Lausanne, CH-1015 Lausanne, Switzerland}
\author{D.B. Tanner}
\affiliation{Department of Physics, University of Florida, Gainesville, Florida 32611, USA}

\date{\today}

\begin{abstract}
We report measurements of Shubnikov-de Haas (SdH) oscillations in single crystals of BiTeCl at magnetic fields up to 31 T and at temperatures as low as 0.4 K. Two oscillation frequencies were resolved at the lowest temperatures, $F_{1}=65  \pm 4$ Tesla and $F_{2}=156 \pm 5$ Tesla. We also measured the infrared optical reflectance $\left(\cal R(\omega)\right)$ and Hall effect; we propose that the two frequencies correspond respectively to the inner and  outer Fermi sheets of the Rashba spin-split bulk conduction band. The bulk carrier concentration was  $n_{e}\approx1\times10^{19}$ cm$^{-3}$ and the effective masses $m_{1}^{*}=0.20 m_{0}$ for the inner and $m_{2}^{*}=0.27 m_{0}$ for the outer sheet. Surprisingly, despite its low effective mass, we found that the amplitude of $F_{2}$ is very rapidly suppressed with increasing temperature, being almost undetectable above $T\approx4$ K. 
\end{abstract}
\maketitle

Manipulating the spin of electrons is of growing interest today, both for technological applications 
(spintronics) and for the realization of new, exotic states (triplet superconductivity; Majorana fermions). Topological insulators (TI) have emerged as promising candidates for the realization of these spin-related phenomena, owing to their edge states with spin protected by time-reversal symmetry\cite{Hasan10}. Subsequently, it has been shown theoretically and discovered experimentally that in the \textit{V-VI-VII} layered compounds BiTeX (with X = Cl, Br, I) the spin-orbit interaction can also lift the spin degeneracy of electrons, in a way similar to the effect of time-reversal symmetry at the surface of topological insulators, but this time in the bulk of non-centrosymmetric semiconductors\cite{Ishizaka11, Ermeev12}. The existence of surface states has also been established in these materials\cite{Ermeev12, Landolt12, Crepaldi12}. Theoretical work predicted that under pressure, BiTeI becomes a topological insulator\cite{Bahramy12}; pressure-dependent optical spectroscopy experiment did confirm this prediction\cite{Xi13, Tran13}. 

Adding to the growing interest in the BiTeX compounds, a recent angle-resolved photoemission (ARPES) study discovered that BiTeCl is a topological insulator at ambient pressure\cite{Chen13}. What makes it unique and exciting for the field is that, unlike all previous TIs where the crystal structure preserves inversion symmetry, BiTeCl is the first example of an inversion antisymmetric topological insulator (IATI). The inversion antisymmetry may give rise to other unusual effects, like a strong bulk polarization, topological magneto-electric effect\cite{Fu07}, or topological superconductivity\cite{Qi10}. 

Motivated by these discoveries, we have measured the in-plane magnetoresistance, Hall effect, and optical reflectance of BiTeCl in order to investigate the electronic properties and Fermi surface of this IATI. Single crystals of BiTeCl were grown and characterized according to Ref.\cite{Arnaud13}. A sample of about 3$\times$2$\times$0.1 mm$^{3}$ was cut from a larger piece and gold wires were attached using silver paint for electrical resistance measurements. The experiment was performed in Cell 9 at the National High Magnetic Field Laboratory. This facility combines a top loading $^{3}$He cryostat, with sample in liquid, and a 32 Tesla resistive magnet. Separate measurements of Hall effect were performed using a commercial PPMS system from Quantum Design. Room temperature optical reflectance data $\left(\cal R(\omega)\right)$, at frequencies between 30 and 32 000 cm$^{-1}$ (4 meV- 4 eV), were taken using a combination of a Bruker 113v Fourier spectrometer and a Zeiss microscope photometer. Then, Kramers-Kronig analysis was used to estimate the optical conductivity $\sigma_1(\omega)$. 

\begin{figure}[htp]
\includegraphics[width=0.4\textwidth]{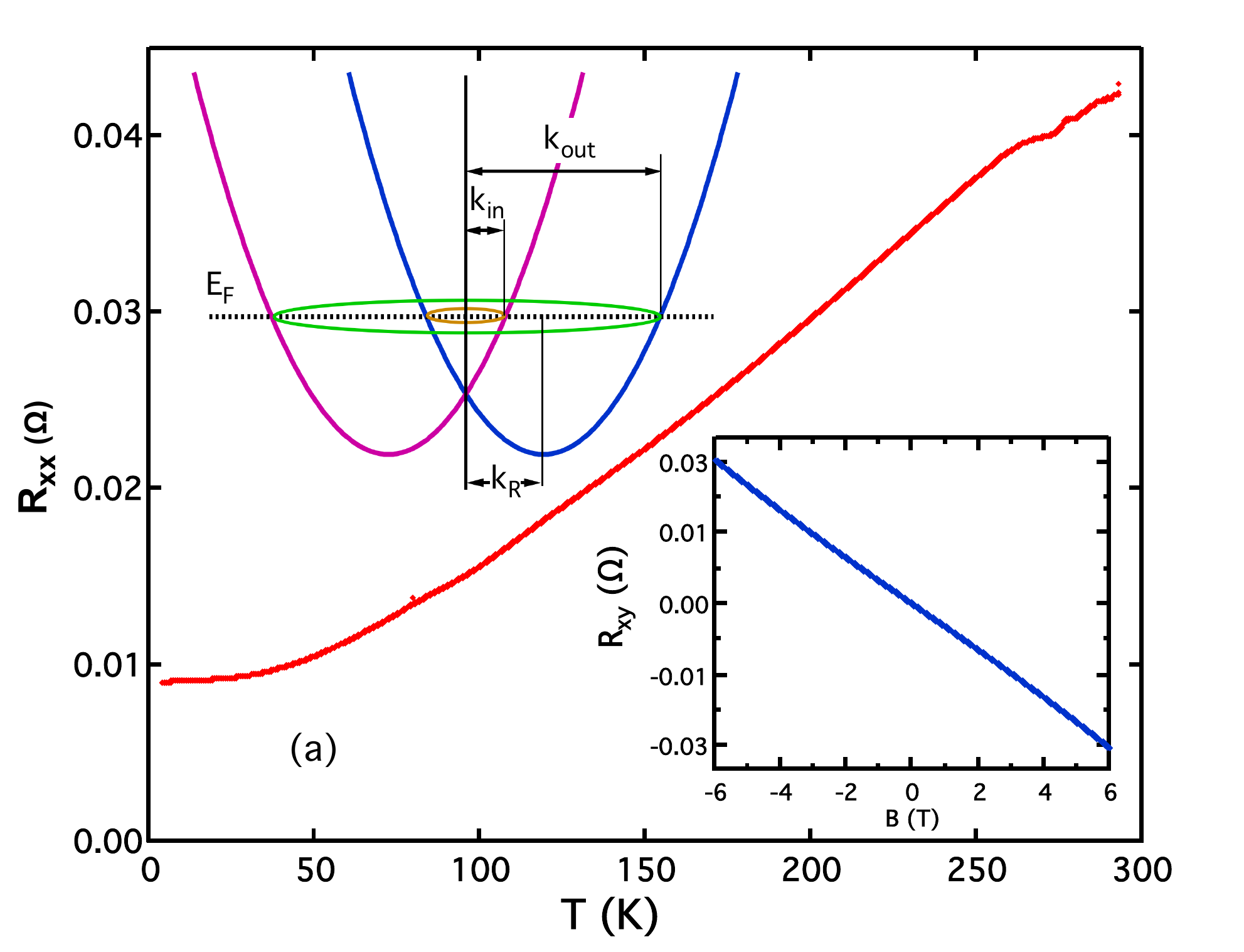}   
\includegraphics[width=0.4\textwidth]{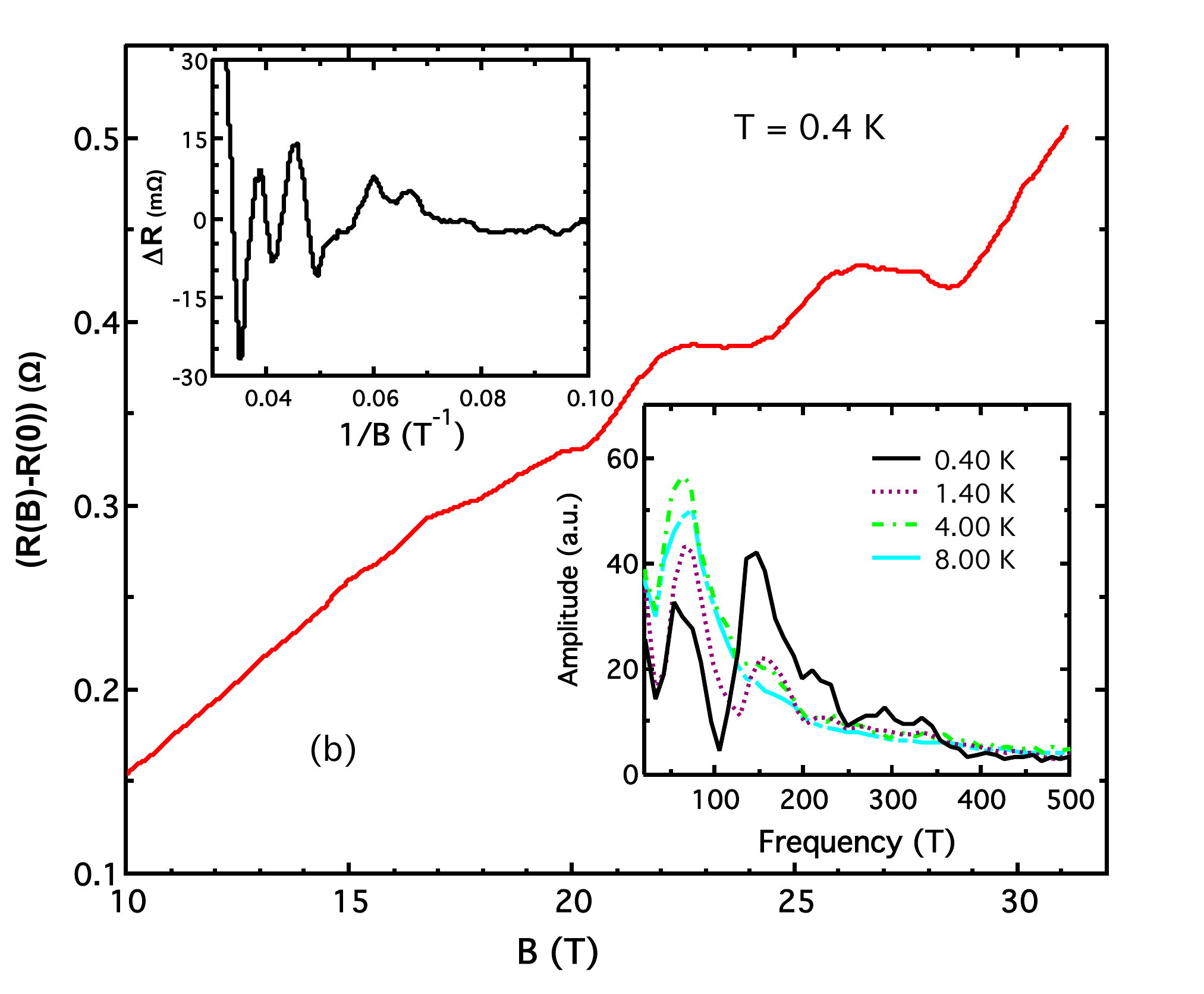}             
\caption{(Color online) (a) Main panel: $R_{xx}$~vs.~$T$ of BiTeCl. Upper inset: Sketch of a Rashba spin-split conduction band showing the momentum ($k_{R}$), the inner and the outer Fermi sheets, of momentum $k_{\rm in}$ and $k_{\rm out}$, respectively. Lower inset: The Hall resistance $R_{xy} (B)$ at $T= 5$~K. (b) Main panel: In-plane magnetoresistance of BiTeCl at high magnetic fields (above 10 T), showing Shubnikov-de Haas oscillations. Upper inset: $\Delta R$ at $T=0.4$~K, obtained after subtracting the continuous background and plotted vs. $1/B$. Lower inset: Fourier transform of $\Delta R(1/B)$ at different temperatures.}  
\label{Fig:Intro}
\end{figure}

The upper panel of Fig.~\ref{Fig:Intro} shows the temperature dependence of the sample resistance. $R_{xx}$ is metallic, with the resistance decreasing by a factor of about 4.5 upon cooling from room temperature to 5K. Hall data, shown in the lower inset of Fig.~\ref{Fig:Intro}(a), shows that the carriers are electrons.  We extracted a carrier concentration $n_{e}\approx 1\times10^{19}$ cm$^{-3}$. Stoichiometric BiTeCl is a semiconductor with energy gap $\approx$220 meV\cite{Chen13}; however, the chemical potential in our sample is clearly into the conduction band, as in most Bi-based semiconductors. 

The lower panel of Fig.~\ref{Fig:Intro} shows the high magnetic field (above 10 T) in-plane magnetoresistance of BiTeCl at $T = 0.4$~K. Oscillatory behavior becomes visible above 15 T. This behavior can be observed more clearly in the upper inset of Fig.~\ref{Fig:Intro}(b), where we plot the data after subtracting a second order polynomial background. First, we notice the non-sinusoidal shape of the oscillations. We suggest that this complex behavior could be due to  beating between two oscillation frequencies. Indeed, despite the observation of only a few oscillations, a Fourier transform of the data in Fig.~\ref{Fig:Intro}(b) reveals two frequencies: $F_{1} =65 \pm4$~T and $F_{2}=156 \pm 5$ T. The corresponding Fermi momentum in the $xy$-plane (using $k_{xy}=\sqrt{2eF/\hbar}$) is $k_{1}=0.04$~\AA$^{-1}$ for $F_{1}$ and $k_{2}$=0.068~\AA$^{-1}$ for $F_{2}$, respectively. We notice that these values for the Fewrmi momentum are much smaller, by a factor of 3 at least, than those reported from ARPES data for the surface states of BiTeCl\cite{Chen13}. Instead, they seem to agree within the uncertainty of the chemical potential level with the results of band structure calculations for the bulk\cite{Ermeev12}. Assuming that the frequencies originate from two 3-D Fermi surfaces, the resulting carrier concentration $n_{3D}=\left(1/3/\pi^2\right)\left(2eF/\hbar\right)^{3/2}$ would be $n_{3D}=3\times 10^{18}$~cm$^{-3}$ for $F_{1}$ and $1\times 10^{19}$~cm$^{-3}$ for $F_{2}$, respectively. We notice that the concentration estimated using $F_{2}$ is nearly identical with that determined above from Hall effect, suggesting a common origin.

\begin{figure}[htp]
\includegraphics[width=0.4\textwidth]{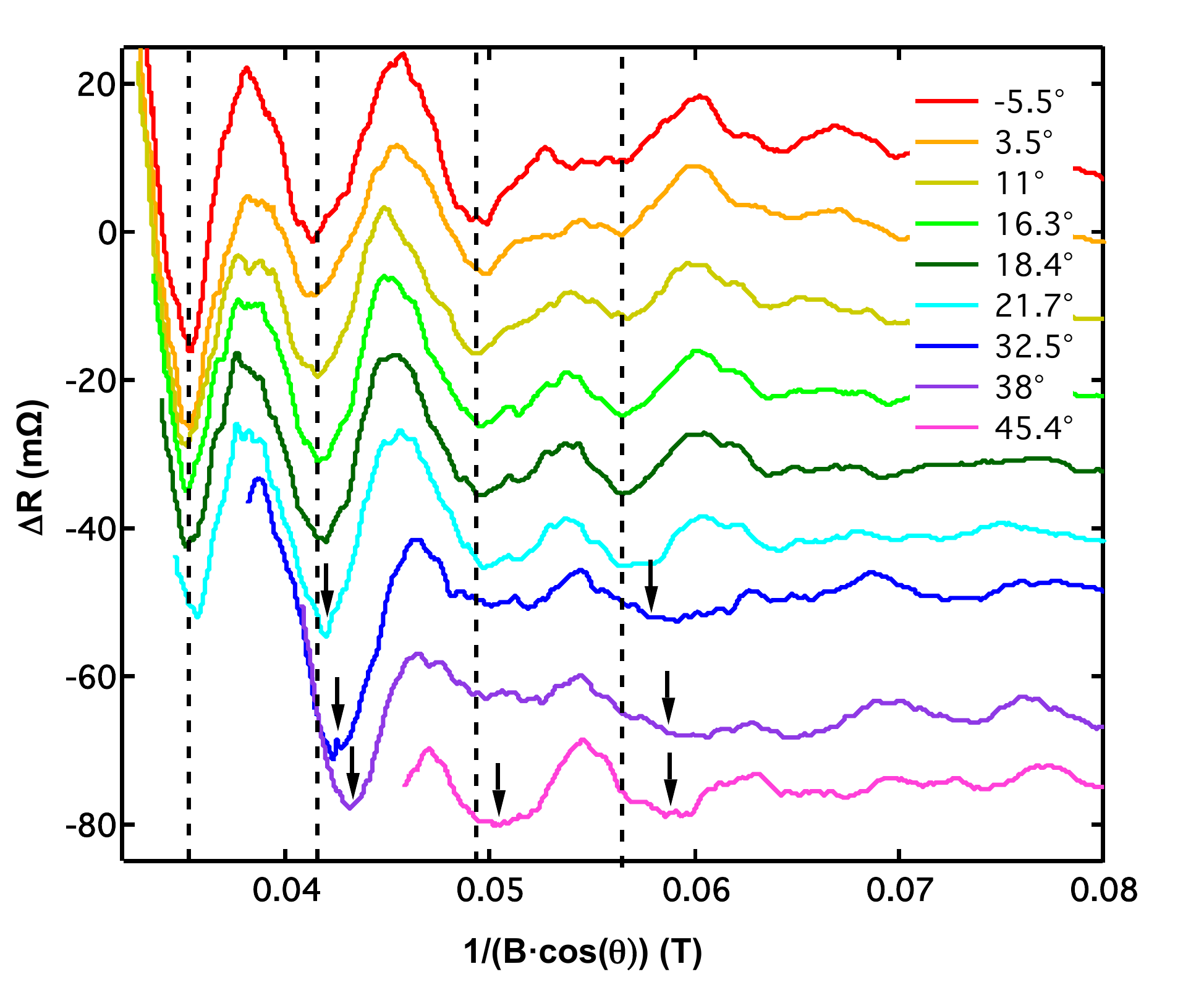}                
\caption{(Color online) $\Delta R(1/B)$ at different angles between the magnetic field and sample surface, plotted against the field component along the normal to sample surface. The dashed line indicates the positions of minima when the field is normal to sample surface and arrows mark the minima at higher angles.}   
\label{Fig:Angle}
\end{figure}

In Figure~\ref{Fig:Angle} we plot the angle dependent SdH oscillations at $T=0.4$ K as a function of the component of field perpendicular to the sample surface. At small angles, $\theta\leq 20^{\circ}$, the oscillations appear to scale with the normal component of the field. However, at higher angles, there is clear departure for the minima marked in the figure. This deiviation is indicative of a three-dimensional Fermi surface;  moreover, it makes it unlikely that the oscillations originate from surface carriers. Were that the case, they would  be highly two-dimensional,   scaling precisely with the normal component of the magnetic field. Furthermore, a strong suppression of amplitude with tilting field is expected for surface oscillations, which does not appear to be the case in our data, at least up to $\theta\approx 45^{\circ}$. Experiments at higher magnetic field and larger tilting angle would help prove our conclusion more clearly. 

\begin{figure}[htp]
\includegraphics[width=0.4\textwidth]{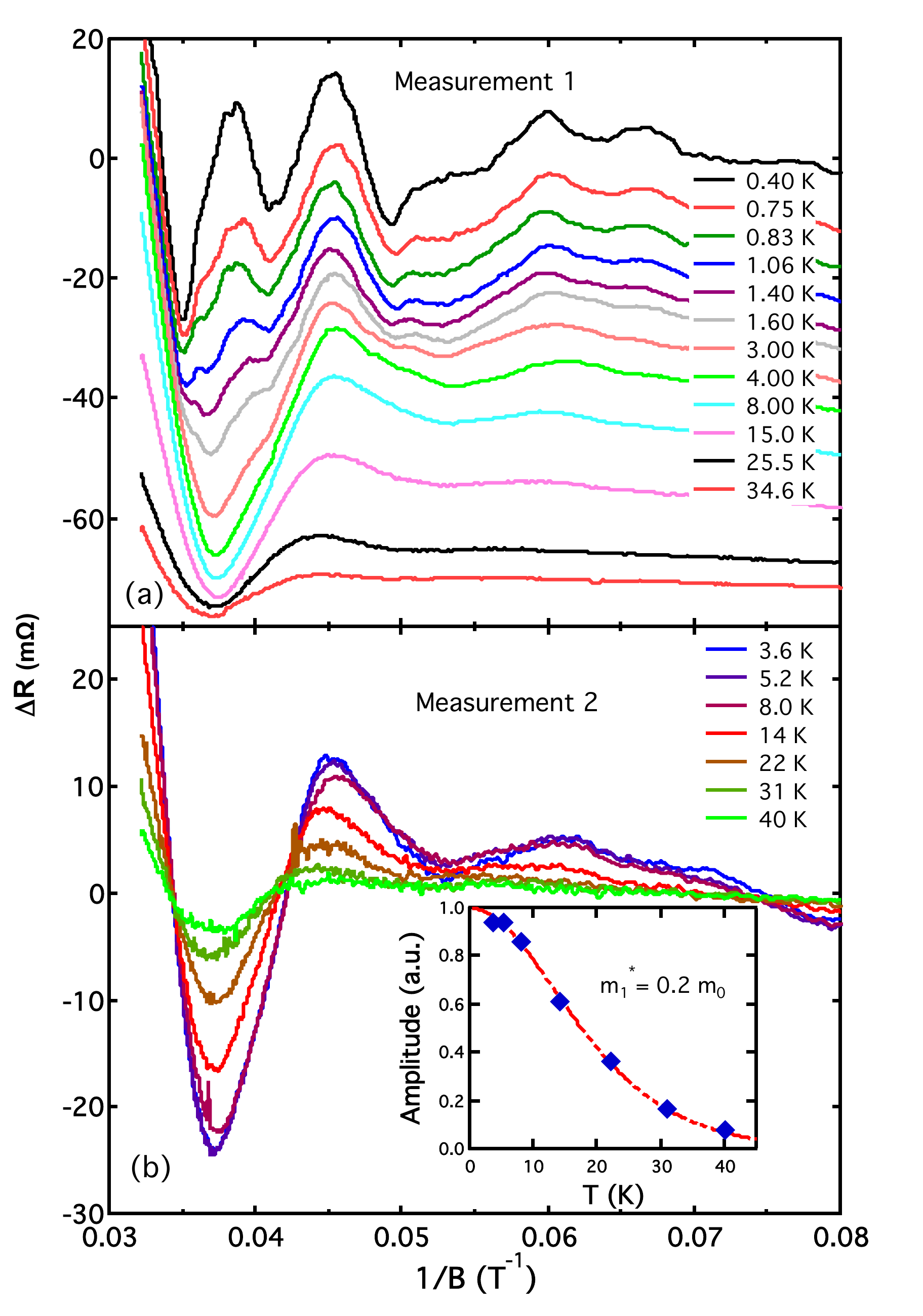}                
\caption{(Color online) $\Delta R$ vs.$1/B$ (with $B$ applied perpendicular to the sample surface) at temperatures from 0.4 to 40~K, for two different measurements, as explained in the main text. Inset of (b): Amplitude of the $F_{1}$ oscillation at $B = 0.037$~T$^{-1}$ (symbols) and a fit to Lifshitz-Kosevich formula.}   
\label{Fig:Temp}
\end{figure}

The temperature dependence of the quantum oscillations in BiTeCl is shown in Fig.~\ref{Fig:Temp}(a). It can be clearly seen that the higher frequency, $F_{2}$, is strongly suppressed with increasing temperature, being almost undetectable above $T\approx4$~K. In contrast,  the low frequency ($F_{1}$) oscillations can be observed up to $T\approx35$~K. The lower inset of Fig.~\ref{Fig:Intro}, displaying the Fourier transform of the SdH oscillations for several temperatures, illustrates the same thing. There is a rapid decrease of the $F_{2}$ amplitude with temperature and an enhancement of $F_{1}$ once $F_{2}$ is suppressed. Such behavior suggests that at low temperature the amplitude of oscillations is affected by the beating between the two frequencies, and once one is strongly damped with increased temperature, the other emerges more clearly. 

We find the strong suppression of the $F_{2}$ amplitude with temperature puzzling. First, whether $F_{2}$ originates from bulk or surface carriers, it would imply a carrier effective mass much larger than previously observed and expected for these semiconductors, where in general  $m^{*} \lesssim m_{0}$, the free-electron mass. Second, repeating the experiment with different locations of the contacts (although still in the plane of the sample) we found the same situation, except that the frequency $F_{2}$ was even more strongly suppressed above 4~K. This suppression can be clearly seen in the main panel of Fig.~\ref{Fig:Temp}(b), where we show the higher temperature (T$\geq$4K) SdH oscillations for the second experiment. It appears that the data contain only one frequency ($F_{1}$), and indeed, we found almost no hint of $F_{2}$ in the Fourier transform. While we do not have yet an explanation for this behavior, the presence of only one frequency, $F_{1}$, allowed us to determine more accurately the temperature dependence of its amplitude and hence, its carrier effective mass $m^{*}_{1}$. In the inset of Fig.~\ref{Fig:Temp}(b) we plot the amplitude of the oscillation with minimum at $B = 0.037$~T$^{-1}$ versus temperature. A fit to the Lifshitz-Kosevich formula\cite{Shoenberg84} yields $m^{*}_{1} = (0.2\pm 0.03) m_{0}$.

\begin{figure}[htp]
\includegraphics[width=0.45\textwidth]{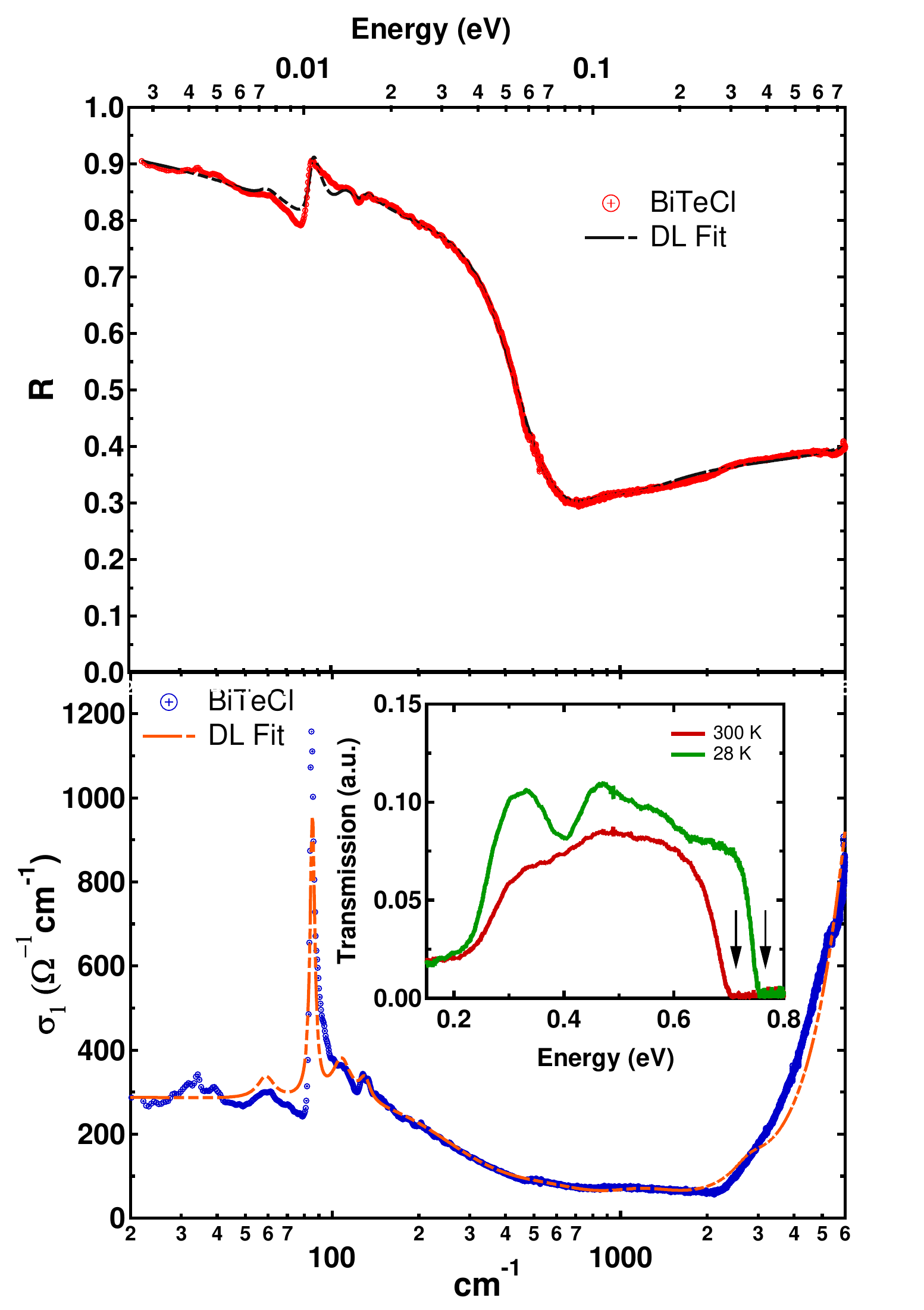}                
\caption{(Color online) (a)~$\cal R(\omega)$ at $T = 300$~K (red symbols) and a Drude-Lorentz fit (dashed dotted black line) of BiTeCl. (b)Main panel: Real part of optical conductivity $\sigma(\omega)$ obtained from Kramers-Kronig analysis (blue symbols) and the Drude-Lorentz fit (dashed dotted orange line). Inset: Optical transmission of sample at room temperature and 28 K. The arrows indicate the onset of the semiconducting gap.}   
\label{Fig:Refl}
\end{figure}

Given that both Hall effect and SdH oscillations may be affected, or even dominated, either  by  bulk or by surface carriers, we measured the infrared  reflectance $\cal R(\omega)$, as a true bulk probe. We showed previously\cite{Martin13} for the case of BiTeI  that a monolayer-thick conducting surface layer  with impedance $R_\square \sim 2000$~$\Omega_\square$  affects the far-infrared reflectance of a conductor with conductivity of order $\sigma_1 \sim  100$~$\Omega^{-1}$cm$^{-1}$  by less than 0.5\%. Therefore,  $\cal R(\omega)$ of BiTeCl, shown in Fig.~\ref{Fig:Refl}(a), should be governed mainly by the response of the bulk carriers. A clear plasma edge can be observed around 500 cm$^{-1}$ and in the limit $\omega\rightarrow$0, $\cal R(\omega)$ exceeds 90\%, consistent with metallic behavior of the bulk. 

Clear phonon modes are observed at low frequency; they will be discussed elsewhere.  Kramers-Kronig analysis of the reflectance yields the optical conductance, $\sigma_1(\omega)$, shown in Fig.~\ref{Fig:Refl}(b). A clear zero frequency (Drude) peak and sharp phonon modes are visible, as well as an absorption edge, at about 250 meV. The estimate of the semiconducting gap from optical reflectance is howevere strongly affected by the fact that the sample also trasmits light in the mid-infrared range, as it can be observed from our transmission measurements in the inset of Fig.~\ref{Fig:Refl}(b). Within the uncertainty of the chemical potential with respect to the conduction band minimum, we find that the semiconducting gap is of the order of 760 meV, consistent with previous optical data~\cite{Akrap14}. We notice also a similar interband transition to that observed and discussed in Ref.~\cite{Akrap14}, situated at slightly higher energy, about 400 meV, in our sample. The dc conductivity is $\sigma(0)\approx$ 300~$\Omega^{-1}cm^{-1}$, i.e. $\rho_{0}\approx  3$ m$\Omega$-cm, characteristic of a moderately doped semiconductor. In order to extract the free carrier properties, we fit both $\cal R(\omega)$ and $\sigma(\omega)$ with a Drude-Lorentz model\cite{Wooten72}, optimizing the fit so that the same set of parameters reproduces best both quantities. The fit  can be seen in both Fig.~\ref{Fig:Refl} (a) and (b). The Drude plasma frequency is $\omega_{p}= \sqrt{n_{e}e^{2}/m^{*}\epsilon_{0}}=1750$~cm$^{-1}$ and the scattering rate $1/\tau  = 180$~cm$^{-1}$. 

On the one hand, if we consider the carrier concentration as determined from the Hall effect, which was identical with that resulting from the larger SdH oscillation frequency $F_{2}$, then $\omega_{p}$ gives for the effective mass of the bulk the value $m^{*}\approx 0.27m_{0}$. On the other hand, if were to assign $\omega_{p}$ to the lower frequency $F_{1}$, then the resulting effective mass would disagree by almost an order of magnitude with that determined above from the Lifshitz-Kosevich formula. Therefore, it is more likely that the higher frequency $F_{2}$ originates from the same bulk conduction band that dominates the optical response. Furthermore, if this is the case, it makes even more unlikely that $F_{1}$ originates from the surface. First, let us recall that its Fermi momentum is much smaller than that measured previously for the surface states. Second, a smaller pocket for the surface than bulk is contradictory to both previous theoretical and experimental findings for the band structure of BiTeCl\cite{Chen13, Ermeev12}.

In conclusion, we have observed   two SdH oscillations  in our magnetoresistance data. Both originate from  bulk Fermi sheets. The most likely scenario for  BiTeCl is that they correspond to the inner and outer Fermi sheets of the Rashba spin-split bulk conduction band, as sketched in the inset in the upper panel of Fig.~\ref{Fig:Intro}. In this case, according to the geometry illustrated in this diagram, we can determine the bulk Rashba momentum $k_{R}$, as $k_{R}=\left(k_{\rm out}-k_{\rm in}\right)/2$, where $k_{\rm out}=k_{2}$, the Fermi momentum of the outer band and $k_{\rm in}=k_{1}$, the momentum of the inner band, respectively. We find $k_{R}=0.012$~\AA$^{-1}$, which suggests that the Rashba momentum in BiTeCl is smaller by a factor of about 4.5 than that of BiTeI. The precise origin for the strong suppression of the amplitude of the higher SdH oscillation with temperature remains unknown. We suggest the possibility of oscillation beating between the two frequencies. Higher magnetic field experiments may help elucidate the question.

A portion of this work took place at the University of Florida,  supported by the DOE through Grant No. DE-FG02-02ER45984.  A portion was performed at the National High Magnetic Field Laboratory, which is supported by National Science Foundation Cooperative Agreement No. DMR-0654118, the State of Florida, and the U.S. Department of Energy. We   thank Bobby Joe Pullum for support with the experiment at the National High Magnetic Field Laboratory. We also thank Dmitrii Maslov for stimulating discussions.

Note: While preparing our manuscript, we became aware of the posting of a similar work. Reference \cite{Xiang14} assigns the SdH oscillations to the surface states. However, here we provide compelling evidence in this work, taken at higher magnetic fields and lower temperatures, that the oscillations originate from the bulk Fermi surface of BiTeCl.

\end{document}